\documentclass[11pt]{article}
\usepackage[margin=1in]{geometry}
\usepackage{mathtools,tensor,graphicx,amsmath,amssymb}
\usepackage{jheppub}

\title{Celestial Current Algebra from Low's Subleading Soft Theorem}
\author[a]{Elizabeth Himwich}
\author[a]{and Andrew Strominger}

\affiliation[a]{Center for the Fundamental Laws of Nature, Harvard University, Cambridge, MA, USA}
\emailAdd{himwich@g.harvard.edu}
\emailAdd{strominger@physics.harvard.edu}

\abstract{The leading soft photon theorem implies that four-dimensional scattering amplitudes are controlled by a  two-dimensional (2D) $U(1)$ Kac-Moody symmetry that acts on the celestial sphere at null infinity ($\mathcal{I}$). This celestial  $U(1)$ current is realized  by components of the electromagnetic vector  potential on the boundaries of $\mathcal{I}$. Here, we develop a parallel story for Low's subleading soft photon theorem. It gives rise to a second celestial current, which  is realized  by  vector potential components that are subleading in the large radius expansion about the boundaries of $\mathcal{I}$. The subleading soft photon theorem is reexpressed as a celestial Ward identity for  this second current, which involves novel shifts by one unit in  the conformal dimension of charged operators.}
\arxivnumber{1901.01622}
\begin{document}
\maketitle


\section{Introduction} 

In any four-dimensional (4D) theory with photons, the soft photon theorem implies \cite{Strominger2014b,He2014,Kapec2015,Cheung2017,Nande2018} the existence of a two-dimensional (2D) $U(1)$ Kac-Moody symmetry.  The consequences  of the symmetry become  most transparent when 4D scattering amplitudes are reexpressed as correlation functions on the celestial sphere  at null infinity ($\mathcal{I}$), on which  the 4D Lorentz group acts as the 2D Euclidean conformal group. The Kac-Moody  currents act on this celestial sphere and are sourced by electromagnetic charge currents that cross it.  All amplitudes are thereby highly constrained, and in particular are set to zero by infrared divergences \cite{Kapec2017b} if the associated conservation laws are violated. The celestial Kac-Moody current may be explicitly realized by a sum of the gauge potentials on the $S^2$  boundaries of $\mathcal{I}$, denoted $A_{\bar{z}}^{(0)}$ below. This story is reviewed in \cite{Strominger2018}.

In the 1950s Low and others \cite{Low1954,GellMann1954,Low1958,Burnett1968,Weinberg2005} established a second, universal, relation governing the subleading term in the soft expansion of an asymptotic photon. A similar 
story is expected to derive from this universal relation, but so far is only partially understood \cite{Lysov2014, Campiglia2016, Conde2017}.

In this paper we show that the subleading soft theorem implies a second current algebra on the celestial sphere. The currents are the constructed
from the boundary values of the subleading term of the gauge potential, denoted $A_{\bar{z}}^{(1)}$, in the large radius expansion around $\mathcal{I}$. Naively, 
$A_{\bar{z}}^{(1)}$ is determined from the leading potential $A_{\bar{z}}^{(0)}$ by the equations of motion and is not an independent field. However, in attempting to explicitly solve for $A_{\bar{z}}^{(1)}$ in terms of  $A_{\bar{z}}^{(0)}$, one encounters an integration function on the sphere. This implies that the boundary values of $A_{\bar{z}}^{(1)}$ are independent fields after all, and in fact turn out to comprise an independent ``subleading'' current algebra.

The current algebra generated on the celestial sphere by boundary values of $A_{\bar{z}}^{(1)}$ has interesting and unconventional features. The OPE of the subleading current with a charged operator with 2D conformal weights  $(h,\bar h)$ shifts the weights  to $(h-{1 \over 2},\bar h-{1 \over 2})$. This is possible because such operators lie in the continuous unitary  principal series. Our main result is formula \eqref{currentalg} below which describes this action. 
It will be interesting to eventually understand the constraints of \eqref{currentalg} on scattering amplitudes. 

In this note we make the simplifying restrictions that there are no longe range magnetic fields near spatial infinity and also that charge is carried by massless scalar fields. 
We expect our results to hold in a more general context, as their form is  largely dictated by symmetries.  

This note is organized as follows. In Section 2, we introduce our conventions and present basic formulas. In Section 3, we rewrite the subleading soft theorem as a relation between the boundary values of a subleading gauge parameter. In Section 4, we take the quantum matrix element of this conservation law  and express it as a Ward identity for a novel 2D current algebra on the celestial sphere.  Appendix A gives some details of the asymptotic expansion about $\cal{I}$ in Lorenz gauge. 

\section{Maxwell Equations in Lorenz Gauge}
We largely employ the retarded (advanced) coordinates on flat Minkowski space
\begin{equation}
ds^2 = -du^2 - 2dudr + 2r^2 \gamma_{z\bar{z}}dz d\bar{z} =  -dv^2 + 2dvdr + 2r^2 \gamma_{z\bar{z}}dz d\bar{z},
\end{equation}
with $u \ (v)$ retarded (advanced) time and $\gamma_{z\bar{z}}=2/(1+z\bar{z})^2$ the unit round metric on $S^2$.  These are related to the Cartesian coordinates $(x^0,x^1,x^2,x^3)$ by 
\begin{equation}
\begin{aligned}
x^0 &= u + r = v-r \\
x^1 + i x^2 &= \frac{2rz}{1 + z\bar{z}} \\
x^3 &= \frac{r(1-z\bar{z})}{1 + z\bar{z}}.
\end{aligned}
\end{equation}
In this paper we use the  the Lorenz gauge condition $\nabla_{\mu}A^{\mu} = 0$. The Maxwell equations $\nabla_{\mu}F^{\mu\nu} = e^2j^{\nu}$ in this gauge in retarded coordinates are
\begin{equation} \label{maxwell}
\begin{aligned}
-2r\partial_r(r\partial_u A_u) + \partial_r(r^2 \partial_r A_u) + 2\gamma^{z\bar{z}}\partial_z\partial_{\bar{z}}A_u &= e^2 r^2 j_u \\
- 2\partial_r(r A_u)  - 2\partial_u\partial_r(r^2 A_r) + \partial_r^2(r^2 A_r) + 2\gamma^{z\bar{z}}\partial_z\partial_{\bar{z}}A_r &= e^2 r^2 j_r\\
-2r^2\partial_u\partial_rA_z + r^2\partial_r^2A_z + 2r\partial_z(A_r - A_u) + 2 \partial_z(\gamma^{z\bar{z}}\partial_{\bar{z}}A_z) &= e^2 r^2 j_z. 
\end{aligned}
\end{equation}
See Appendix A for further details.

\section{Subleading Soft Theorem as Subleading Gauge Transformation}

Low's subleading soft photon theorem, following the notation of~\cite{Lysov2014}, can be written as an asymptotic symmetry acting on in- and out-states. Denote a state with $n$ massless hard particles of energies $\omega_k$, charges $eQ_k$ and momenta
\begin{equation}
  p^{\mu}_k = \frac{\omega_k}{1 + z_k\bar{z}_k}(1 + z_k\bar{z}_k, z_k + \bar{z}_k, i(\bar{z}_k - z_k), 1 - z_k \bar{z}_k)
\end{equation}
by $|z_1, \ldots \rangle$ and hard $S$-matrix elements by $\langle z_{n+1}, \ldots | \mathcal{S} | z_1, \ldots \rangle$. The Low-Burnett-Kroll-Goldberger-Gell-Mann soft theorem says that if a positive helicity photon with energy $\omega \rightarrow 0$, the first two terms of the soft expansion are
\begin{equation}
  \langle z_{n+1}, \ldots | a_{-}^{out}(\vec{q})\mathcal{S} | z_1, \ldots \rangle = (J^{(0)-} + J^{(1)-}) \langle z_{n+1}, \ldots | \mathcal{S} | z_1, \ldots \rangle,
\end{equation}
with
\begin{equation}
  J^{(0)-} = e \sum_k Q_k \frac{p_k \cdot \varepsilon^{-}}{p_k \cdot q} \sim \mathcal{O}(\omega^{-1}), \ \ J^{(1)-} = -i e \sum_k Q_k \frac{q_{\mu} \varepsilon_{\nu}^{-}J_k^{\mu\nu}}{p_k \cdot q}  \sim \mathcal{O}(\omega^{(0)})
\end{equation}
with $\varepsilon$ the photon polarization and $J_k^{\mu \nu}$ the total angular momentum of the $k^{\text{th}}$ particle. The $J^{(0)-}$ contribution can be eliminated with the projection operator $(1 + \omega \partial_{\omega})$. In~\cite{Lysov2014} it was shown that, for the special case of a scalar field with $J_{k\mu\nu} = -i \left( p_{k\mu}\frac{\partial}{\partial p_k^{\nu}} -  p_{k\nu}\frac{\partial}{\partial p_k^{\mu}}\right)$, rewriting $(p_k^{\mu},q^{\mu})$ in terms of $(E_k, z_k, \bar{z}_k)$ gives
\begin{equation} \label{softthm}
  \begin{aligned}
    \lim_{\omega \to 0}&(1 + \omega\partial_{\omega}) \langle z_{n+1}, \ldots | a_{-}^{out}(\vec{q})\mathcal{S} | z_1, \ldots \rangle \\
    &= - e \sum_k \frac{Q_k}{\sqrt{2}(\bar{z}_k - \bar{z})} \big[(1 + z\bar{z}_k)) \partial_{E_k} + \frac{1}{E_k}(z-z_k)(1 + z_k\bar{z}_k) \partial_{z_k} \big] \langle z_{n+1}, \ldots | \mathcal{S} | z_1, \ldots \rangle. 
  \end{aligned}
\end{equation}
As in~\cite{Lysov2014} it is useful to define operators $F_{u\bar{z}}^{sub}$ that create subleading soft photons. These are defined on $\mathcal{I}^+$ in terms of the photon polarization
\begin{equation}
  \hat{\varepsilon}^+_{\bar{z}} = \frac{\partial_{\bar{z}}x^{\mu}}{r}\varepsilon^+_{\mu} = \frac{\sqrt{2}}{1 + z\bar{z}}
\end{equation}
by
\begin{equation} \label{Fsub}
  \begin{aligned}
    F_{u\bar{z}}^{sub} :=& \int du u F_{u \bar{z}}^{(0)} = \int du u \partial_u A_{\bar{z}}^{(0)} \\
    =& \frac{ie}{8\pi}\hat{\varepsilon}^+_{\bar{z}}\lim_{\omega \to 0}(1 + \omega \partial_{\omega})[ a_{-}^{out}(\omega \hat{x}) - a_+^{out}(\omega \hat{x})^{\dagger}].
  \end{aligned}
\end{equation}
The fields in this expression, and subsequent expressions, are the functions of $(u,z,\bar z)$ that appear as coefficients in the asymptotic ${1 \over r}$ expansion about $\mathcal{I}^+$. The order   $\frac{1}{r^n}$ at which they appears in this expansion  is  denoted by the superscript $(n)$.  For simplicity, we restrict here to the case where there are no long range magnetic fields near spatial infinity so that $A_{{z}}^{(0)}$ is pure gauge and $F_{z\bar{z}}^{(0)}=0$ at $\mathcal{I}^+_\pm$. Multiplying \eqref{softthm} by $\hat{\varepsilon}^+_{\bar{z}}$ and then acting with $\gamma^{z \bar{z}}D_{\bar{z}}D_z$ and using \eqref{Fsub} gives 
\begin{equation} \label{action}
  \begin{aligned}
    \langle z_{n+1}, \ldots &| Q_S^+ \ \mathcal{S} | z_1, \ldots \rangle \\
    &= - \frac{i}{4 \pi} \sum_k Q_k \big[ - 2\pi \gamma^{z \bar{z}}D_{\bar{z}} \delta^2( z - z_k)\partial_{E_k} + \frac{1}{E_k} \frac{\gamma^{z_k \bar{z}_k}}{(\bar{z} - \bar{z}_k)^2} \partial_{z_k} \big] \langle z_{n+1}, \ldots | \mathcal{S} | z_1, \ldots \rangle,
  \end{aligned}
\end{equation}
where we can define the ``soft'' charge
\begin{equation} \label{softcharge}
  Q_S^+ = \frac{2}{e^2}\int du \  u \partial_u \gamma^{z\bar{z}}D_{\bar{z}}D_{z} A_{\bar{z}}^{(0)}.
\end{equation}
For the leading soft charge, the analog of the soft term is a total $u$-derivative and reduces to a difference between two terms on the boundaries of $\mathcal{I}^+$, signalling the central role  of $\mathcal{I}$ boundary dynamics. In contrast, this total derivative structure is not manifest in the soft term given in~\cite{Lysov2014} and in \eqref{softcharge}. 
However, we now show that this structure reappears when $\mathcal{Q}^+$ is reexpressed in terms of the subleading component $A_z^{(1)}$ of the gauge field, which enables one to rewrite it in terms of hard currents and the $\mathcal{I}^+_\pm$ boundary values of $A_z^{(1)}$. The elimination of $A_{\bar{z}}^{(0)}$ from \eqref{softcharge}  in favor of $A_{\bar{z}}^{(1)}$ proceeds via the asymptotic expansion of the Maxwell equations, which are without sources for the soft insertion (see Appendix A for details) 
\begin{equation} \label{eq:Maxwell}
 \partial_u^2A_{\bar{z}}^{(1)} = -  \partial_uD^zD_zA_{\bar{z}}^{(0)}. 
\end{equation}
This allows us to rewrite the soft charge as
\begin{equation}
  Q_S^+ = \frac{2}{e^2} ( 1 - u \partial_u) A_{\bar{z}}^{(1)}  \Bigg{|}^{\mathcal{I}^+_+}_{\mathcal{I}^+_-}. 
\end{equation}
Lorenz gauge $\nabla^\mu A_\mu=0$ leaves unfixed residual gauge transformations of the form $A_{\mu} \rightarrow A_{\mu} + \partial_{\mu}\varepsilon$ with  $\square \varepsilon = 0$. The solution to this equation in retarded coordinates requires two pieces of free data, at different orders in the asymptotic expansion: the free function $\varepsilon^{(0)}(z,\bar{z})$, which is related to the leading soft theorem, and the free function $\varepsilon^{(1)}(u,z,\bar{z})$, which is independent free data. This latter residual freedom enables  us to fix the  subsidiary gauge condition 
\begin{equation} A_u^{(1)} = 0,\end{equation}
which implies that $\partial_u\varepsilon^{(1)} = 0$. We are left with a free function $\varepsilon^{(1)}(z,\bar{z})$. The gauge transformations are parametrized as
\begin{equation}
\varepsilon=\varepsilon^{(0)}(z,\bar{z})+ \frac{u}{2}D^2\varepsilon^{(0)}(z,\bar{z}) \frac{\log r}{r} + \frac{\varepsilon^{(1)}(z,\bar{z})}{r}+...
\end{equation}
At early and late times along future null infinity, where the matter current is zero, the field configurations return to pure gauge. Hence the asymptotic behavior near $\mathcal{I}^+_\pm$ is
\begin{equation} \label{req}
A_{\bar{z}\pm}^{(0)}= D_{\bar{z}}\varphi^{(0)}_{\pm}(z,\bar{z}),~~~\tilde{A}_{\bar{z}\pm}^{(1)} = \frac{u}{2} D_{\bar{z}} D^2 \varphi^{(0)}_{\pm}(z,\bar{z}),~~~A_{\bar{z}\pm}^{(1)}= D_{\bar{z}}\varphi^{(1)}_{\pm}(z,\bar{z}),
\end{equation}
where the tilde denotes a $\log r$ dependence (see Appendix A for details) and where the boundary fields $\varphi$ shift under gauge transformations as   $\varphi^{(0)}_{\pm}\to\varphi^{(0)}_{\pm}+\varepsilon^{(0)}$ and $\varphi^{(1)}_{\pm}\to\varphi^{(1)}_{\pm} + \varepsilon^{(1)}$. The difference in their values at $\mathcal{I}^+_+$ and $\mathcal{I}^+_-$ is determined by the action of the soft factor  and cannot be gauge-fixed to zero. 
To underscore this, we rewrite \eqref{action} as
\begin{equation} \label{action+}
  \begin{aligned}
    \langle z_{n+1}, \ldots | &\left(\frac{2}{e^2} D_{\bar{z}}\varphi^{(1)} \Big{|}^{\mathcal{I}^+_+}_{\mathcal{I}^+_-}\right)\mathcal{S} | z_1, \ldots \rangle \\
    &= - \frac{i}{4 \pi} \sum_k Q_k \big[- 2\pi \gamma^{z \bar{z}}D_{\bar{z}} \delta^2( z - z_k)\partial_{E_k} + \frac{1}{E_k} \frac{\gamma^{z_k \bar{z}_k}}{(\bar{z} - \bar{z}_k)^2} \partial_{z_k} \big] \langle z_{n+1}, \ldots | \mathcal{S} | z_1, \ldots \rangle.
  \end{aligned}
\end{equation}
Similarly, for the insertion of an incoming soft photon $\mathcal{I}^-$,
\begin{equation} \label{action-}
  \begin{aligned}
    \langle z_{n+1}, \ldots | \mathcal{S}  &\left(\frac{2}{e^2} D_{\bar{z}}\varphi^{(1)} \Big{|}^{\mathcal{I}^-_+}_{\mathcal{I}^-_-} \right) | z_1, \ldots \rangle \\
    &=  \frac{i}{4 \pi} \sum_k Q_k \big[- 2\pi \gamma^{z \bar{z}}D_{\bar{z}} \delta^2( z - z_k)\partial_{E_k} + \frac{1}{E_k} \frac{\gamma^{z_k \bar{z}_k}}{(\bar{z} - \bar{z}_k)^2} \partial_{z_k} \big] \langle z_{n+1}, \ldots | \mathcal{S} | z_1, \ldots \rangle.
  \end{aligned}
\end{equation}
To write a shift along all of $\mathcal{I}$, we consider
\begin{equation}
  \begin{aligned}
    \langle z_{n}, \ldots &| Q_S^+  \mathcal{S}  -   \mathcal{S}   Q_S^- |z_1,\ldots \rangle \\
    &= - \frac{i}{2 \pi} \sum_k Q_k \big[ - 2\pi \gamma^{z \bar{z}}D_{\bar{z}} \delta^2( z - z_k)\partial_{E_k} + \frac{1}{E_k} \frac{\gamma^{z_k \bar{z}_k}}{(\bar{z} - \bar{z}_k)^2} \partial_{z_k} \big] \langle z_{n+1}, \ldots | \mathcal{S} | z_1, \ldots \rangle.
    \end{aligned}
\end{equation}
We see that, if there is a nontrivial scattering process, it is impossible to set $A_z^{(1)}=D_z\varphi^{(1)}$ to zero on all boundaries of $\mathcal{I}$, just as the leading soft theorem makes it impossible to set $A_z^{(0)}=D_z\varphi^{(0)}$
to zero on all boundaries.  Hence $\varepsilon^{(1)}$, as well as $\varepsilon^{(0)}$, is a large gauge transformation, and maps one vacuum to a physically inequivalent one. 

\section{Celestial Current Ward Identity}

It is illuminating to rewrite scattering amplitudes  as correlation functions on the celestial sphere, adopting the compact notation \cite{Strominger2018}
\begin{equation}
  \langle z_{n+1}, \ldots | \mathcal{S} | z_1, \ldots \rangle \rightarrow \langle  \mathcal{O}_{E_1}^{(1)}(z_1,\bar{z}_1) \cdots  \mathcal{O}_{E_n}^{(1)}(z_n \bar{z}_n) \rangle.
\end{equation}
In this context, we define the subleading soft photon current
\begin{equation}\label{pik} J_{\bar{z}}^{(1)}= \frac{4\pi}{e^2} \left(D_{\bar{z}}\varphi^{(1)} \Big{|}_{\mathcal{I}^+_+}-2 D_{\bar{z}} \varphi^{(1)} \Big{|}_{\mathcal{I}^+_-}+ D_z\varphi^{(1)} \Big{|}_{\mathcal{I}^-_-} \right),
\end{equation}
where we have used the antipodal matching
\begin{equation}
  \varphi^{(1)} \Big{|}_{\mathcal{I}^+_-}=\varphi^{(1)} \Big{|}_{\mathcal{I}^-_+}. 
\end{equation}
The subleading soft theorem then becomes \footnote{Up to contact terms which will vanish after contour integration.}${}^{,}$\footnote{Using $  - i E\Phi_{kE}^{(2)} = \gamma^{z\bar{z}}D_{\bar{z}} D_{z}\Phi_{kE}^{(1)}$, the right hand side can be rewritten as $\frac{- Q_k \Phi_{kE}^{(2)}(z_k,\bar{z}_k)}{(\bar{z} - \bar{z}_k)} $. This suggests a connection with the identification in ~\cite{Campiglia2016} of subleading soft symmetries with gauge transformations that diverge linearly with $r$. It would be interesting to understand this better.}
\begin{equation} \label{phi1}
  \langle J_{\bar{z}}^{(1)}  \mathcal{O}_{E_1}^{(1)}(z_1,\bar{z}_1) \cdots \mathcal{O}_{E_n}^{(1)}(z_n \bar{z}_n) \rangle = \sum_{k=1}^{n} \frac{- i Q_k }{E_k (\bar{z} - \bar{z}_k)^2} D^{\bar{z}_k} \langle  \mathcal{O}_{E_1}^{(1)}(z_1,\bar{z}_1) \cdots  \mathcal{O}_{E_n}^{(1)}(z_n, \bar{z}_n)\rangle.
\end{equation}
The Mellin transform to a conformal basis for particles with helicity $s$ with conformal weights
\begin{equation}
  (h, \bar{h}) = \frac{1}{2}(\Delta + s, \Delta - s) = \frac{1}{2}(- E\partial_E+ s,-E\partial_E - s)
\end{equation}
is simply
\begin{equation}
  \mathcal{O}_{(h,\bar{h})}(z,\bar{z}) = \int dE \ E^{\Delta - 1}  \mathcal{O}_{E}^{(1)}(z,\bar{z}).
\end{equation}
In this conformal basis, \eqref{phi1} becomes the current algebra relation 
\begin{equation} \label{currentalg}
    \langle  J_{\bar{z}}^{(1)} \mathcal{O}_{(h_1,\bar{h}_1)} \cdots \mathcal{O}_{(h_{n},\bar{h}_{n})} \rangle = -i  \sum_{k } {Q_k \over (\bar{z}-\bar{z}_k)^2 }D^{\bar{z}_k}\langle  \mathcal{O}_{(h_1,\bar{h}_1)} \cdots \mathcal{O}_{(h_k-\frac{1}{2},\bar{h}_k - \frac{1}{2})} \cdots \mathcal{O}_{(h_{n},\bar{h}_{n})} \rangle.
\end{equation}
This is  the celestial representation of the  subleading soft theorem. 

The operators $\mathcal{O}$  which create spacetime particles in a conformal basis appearing in celestial amplitudes are in different types of representations - typically the continuous unitary principal series - than those we are accustomed to in standard 2D CFT. The corresponding amplitudes take a rather different form often involving delta functions on the sphere ~\cite{Pasterski2017a, Pasterski2017b,Schreiber2018,Stieberger2018a}, which makes possible  relations between amplitudes with shifted conformal weights. Relations of this general type were noted in the gravitational context  in \cite{Donnay2018}  and verified  by Stieberger and Taylor~\cite{Stieberger2018b} in some special cases. It would be of interest to 
examine  \eqref{currentalg} in explicit examples. 

Finally, we note that integrating around a contour $C$  weighted by an antiholomorphic function $\varepsilon(\bar{z})$, the subleading soft theorem takes the alternate form 
\begin{equation} \label{coint2}
  \begin{aligned}
    \langle \oint_C \frac{d\bar{z}}{2\pi i} \varepsilon(\bar{z})J_{\bar{z}}^{(1)} \mathcal{O}_{(h_1,\bar{h}_1)} \cdots &\mathcal{O}_{(h_{n},\bar{h}_{n})} \rangle \\
    &= - i \sum_{k \in C} Q_k  \langle  \mathcal{O}_{(h_1,\bar{h}_1)} \cdots \partial_{\bar{z}_k}\varepsilon(\bar{z}_k) D^{\bar{z}_k} \mathcal{O}_{(h_k-\frac{1}{2},\bar{h}_k - \frac{1}{2})} \cdots \mathcal{O}_{(h_{n},\bar{h}_{n})} \rangle,
  \end{aligned}
\end{equation}
where the sum is restricted to operators inside the contour. 

~\nocite{Kapec2015,Donnay2018}

\section*{Acknowledgements}

This work was funded partially by DOE grant DE-SC0007870. E.H. is funded by NSF grant 1745303.  We are grateful to Laura Donnay, Slava Lysov, and Dan Kapec for useful correspondence, and to Monica Pate, Ana Raclariu, and Sabrina Pasterski for discussion and advice throughout this work. 

\appendix

\section{Asymptotic Expansion}
This appendix gives a few details of the large $r$ expansion about $\mathcal{I^+}$. \\

A massless scalar field has $\frac{1}{r}$ expansion near $\mathcal{I}^+$ as
\begin{equation}
\Phi(u,r,z,\bar{z}) =  \sum_{n=1}^{\infty} \frac{\Phi^{(n)}(u,z,\bar{z})}{r^n}.
\end{equation}
The matter currents
\begin{equation}
j_{\mu}=iQ( \bar{\Phi} \partial \Phi - \Phi \partial_{\mu} \bar{\Phi})
\end{equation}
fall off as
\begin{equation}
\begin{aligned}
j_u &\sim \mathcal{O}\left(\frac{1}{r^2}\right), \ \ j_z, j_{\bar{z}} &\sim \mathcal{O}\left(\frac{1}{r^2}\right), \ \ j_r &\sim \mathcal{O}\left(\frac{1}{r^3}\right) .
\end{aligned}
\end{equation}
Finite energy flux and charge  suggest the  falloffs  
\begin{equation}
\begin{aligned}
A_u &\sim \mathcal{O}\left(\frac{1}{r}\right), \ \ A_z, A_{\bar{z}} &\sim \mathcal{O}\left(1\right), \ \ A_r &\sim \mathcal{O}\left(\frac{1}{r^2}\right) .
\end{aligned}
\end{equation}

In order to consistently solve the Maxwell equations in $\nabla^\mu A_\mu=0$ gauge we must allow  logarithmic falloffs in the gauge fields. This gives the expansion
\begin{equation}
\begin{aligned} 
A_{u} &= \sum_{n=2}^{\infty} \frac{A_u^{(n)}}{r^n} + \sum_{m=1}^{\infty} \frac{\tilde{A}_u^{(m)}}{r^m}\log r \\
A_{r} &= \sum_{n=2}^{\infty} \frac{A_r^{(n)}}{r^n} + \sum_{m=2}^{\infty} \frac{\tilde{A}_r^{(m)}}{r^m}\log r \\
A_{z} &= \sum_{n=0}^{\infty} \frac{A_z^{(n)}}{r^n} + \sum_{m=1}^{\infty} \frac{\tilde{A}_z^{(m)}}{r^m}\log r \\
A_{\bar{z}} &= \sum_{n=0}^{\infty} \frac{A_{\bar{z}}^{(n)}}{r^n} + \sum_{m=1}^{\infty} \frac{\tilde{A}_{\bar{z}}^{(m)}}{r^m}\log r . 
\end{aligned}
\end{equation}
Our gauge condition leaves unfixed gauge transformations of the form $\square \varepsilon = 0$, among which are residual gauge transformations with falloff $\mathcal{O}(r^{-1})$ which, like a radiative massless scalar field, have an arbitrary boundary dependence. We have used this freedom to set $A_u^{(1)} = 0$. \\

The Maxwell equations $\nabla_{\mu}F^{\mu\nu} = e^2j^{\nu}$ in retarded coordinates, with $F_{\mu\nu} = \partial_{\mu}A_{\nu} - \partial_{\nu} A_{\mu}$, are
\begin{equation}
\begin{aligned}
(\partial_u - \partial_r)(r^2F_{ru}) + \gamma^{z\bar{z}}(\partial_{\bar{z}}F_{uz} + \partial_{z}F_{u\bar{z}}) &= e^2 r^2 j_u \\
- \partial_r(r^2F_{ru}) + \gamma^{z\bar{z}}(\partial_{\bar{z}}F_{rz} + \partial_{z}F_{r\bar{z}}) &= e^2 r^2 j_r \\
r^2(\partial_r - \partial_u)F_{rz} - r^2 \partial_rF_{uz} - \partial_z(\gamma^{z\bar{z}}F_{z\bar{z}}) &= e^2 r^2 j_{z} \\
r^2(\partial_r - \partial_u)F_{r\bar{z}} - r^2 \partial_rF_{u\bar{z}} - \partial_{\bar{z}}(\gamma^{z\bar{z}}F_{\bar{z}z}) &=  e^2 r^2 j_{\bar{z}} , 
\end{aligned}
\end{equation}
while the Lorenz gauge condition reads 
\begin{equation}
 - \partial_u (r^2 A_r) - \partial_r(r^2 A_u - r^2 A_r) + \gamma^{z \bar{z}}(\partial_z A_{\bar{z}} + \partial_{\bar{z}} A_z) = 0 \ \ . 
\end{equation}
Together these imply
\begin{align} 
\mathcal{O}(\log r):& \nonumber \\
&2\partial_u\tilde{A}_{\bar{z}}^{(1)} - 2\partial_{\bar{z}}\tilde{A}_u^{(1)} = 0 \\
\mathcal{O}(1):& \nonumber \\
&-2\partial_u \tilde{A}_u^{(1)} = e^2 j_u^{(2)} \\
& 2\partial_uA_{\bar{z}}^{(1)} - 2\partial_u\tilde{A}_{\bar{z}}^{(1)} + 2D^zD_zA_{\bar{z}}^{(0)} = e^2 j_{\bar{z}}^{(2)} \ \ . 
\end{align}
where we have used that $\mathcal{O}(\log r)$ expression for $\tilde{j}_{\bar{z}}^{(2)}$ is set to zero because the currents should not have logarithmic falloff. Note that $j_u^{(2)}$ would be incorrectly set to zero if log terms were not included in the expansion. We use these equations to arrive at (\ref{eq:Maxwell}).

\bibliography{../StromingerBib}
\bibliographystyle{JHEP}

\end{document}